\documentclass[11pt]{article}
\usepackage{amssymb,amsmath,amsfonts}
\usepackage{graphicx}
\usepackage{graphics}
\usepackage{eepic,epsfig}

\begin{document}

\title{A summation formula over the zeros of a combination of the associated
Legendre functions with a physical application}
\author{A. A. Saharian\thanks{%
E-mail: saharian@ictp.it } \\
\textit{Department of Physics, Yerevan State University, }\\
\textit{1 Alex Manoogian Street, 0025 Yerevan, Armenia }}
\maketitle

\begin{abstract}
By using the generalized Abel-Plana formula, we derive a summation
formula for the series over the zeros of a combination of the
associated Legendre functions with respect to the degree. The
summation formula for the series over the zeros of the combination
of the Bessel functions, previously discussed in the literature,
is obtained as a limiting case. As an application we evaluate the
Wightman function for a scalar field with general curvature
coupling parameter in the region between concentric spherical
shells on background of constant negative curvature space. For the
Dirichlet boundary conditions the corresponding mode-sum contains
series over the zeros of the combination of the associated
Legendre functions. The application of the summation formula
allows us to present the Wightman function in the form of the sum
of two integrals. The first one corresponds to the Wightman
function for the geometry of a single spherical shell and the
second one is induced by the presence of the second shell. The
boundary-induced part in the vacuum expectation value of the field
squared is investigated. For points away from the boundaries the
corresponding renormalization procedure is reduced to that for the
boundary-free part.
\end{abstract}

\bigskip


\bigskip

\section{Introduction}

\label{sec:Introd}

The associated Legendre functions are an important class of special
functions that appear in a wide range of problems of mathematical physics.
The physical importance of these functions is related to the fact that they
appear as solutions of the field theory equations in various situations. In
particular, the radial parts of the solutions for the scalar, fermionic and
electromagnetic wave equations on background of constant curvature
spacetimes are expressed in terms of the associated Legendre functions (see,
for instance, \cite{Grib94,Most97,Birr82}). The eigenfunctions in braneworld
models with de Sitter and anti-de Sitter branes are also expressed in terms
of these functions (see \cite{Noji00}). Motivated by this, in \cite{Saha08},
by making use of the generalized Abel-Plana formula, we have derived a
summation formula for the series over the zeros of the associated Legendre
function of the first kind with respect to the degree (for the generalized
Abel-Plana formula and its applications to physical problems see \cite%
{Sah1,Saha00Rev,Saha07Rev}). This type of series is contained in
the mode-sum for two-point functions of a quantum scalar field in
background of a constant curvature space with spherical boundary,
on which the field obeys the Dirichlet boundary condition. The
application of the summation formula allowed us to extract from
the vacuum expectation values the part corresponding to the
situation without boundary and to present the boundary-induced
part in terms of rapidly convergent integral.

In the corresponding problem with two concentric spherical
boundaries, in the region between two spheres the eigenfunctions
are the combination of the associated Legendre functions of the
first and second kinds. The eigenfrequences are determined by the
location of the zeros of this combination with respect to the
degree. In the present paper, by specifying the functions in the
generalized Abel-Plana formula, we obtain a summation formula for
the series over these zeros. As in the case of the other
Abel--Plana-type formulae, previously considered in the
literature, this formula presents the sum of the series over the
zeros of the combination of the associated Legendre function in
the form of the sum of two integrals. In boundary-value problems
with two boundaries the first integral corresponds to the
situation when one of the boundaries is absent and the second one
presents the part induced by the second boundary. For a large
class of functions the latter is rapidly convergent and, in
particular, is useful for the numerical evaluations of the
corresponding physical characteristics.

The paper is organized as follows. In section \ref{sec:SumForm}, by
specifying the functions in the generalized Abel-Plana formula we derive a
formula for the summation of series over zeros of the combination of the
associated Legendre functions with respect to the degree. In section \ref%
{sec:Special}, special cases of this summation formula are considered.
First, as a partial check we show that as a special case the standard
Abel-Plana formula is obtained. Then we show that from the summation formula
discussed in section \ref{sec:SumForm}, as a limiting case the formula is
obtained for the summation of the series over the zeros of the combinations
of the Bessel functions, previously derived in \cite{Sah1}. A physical
application is given in section \ref{sec:Phys}, where the positive frequency
Wightman function for a scalar field is evaluated in the region between two
spherical boundaries on background of a negative constant curvature space.
It is assumed that the field obeys Dirichlet boundary condition on the
spherical shells. The use of the summation formula from section \ref%
{sec:SumForm} allows us to extract from the vacuum expectation value the
part corresponding to the geometry where the outer sphere is absent. The
part induced by the latter is presented in terms of an integral, which is
rapidly convergent in the coincidence limit for points away from the sphere.
The main results of the paper are summarized in section \ref{sec:Conclus}.
In appendix \ref{sec:Zeros} the formula for the normalization integral is
derived and we show that the zeros of the combination of the associated
Legendre functions with respect to the degree are simple.

\section{Summation formula}

\label{sec:SumForm}

Let $z=z_{k}$, $k=1,2,\ldots $, be zeros of the function
\begin{equation}
X_{iz}^{\mu }(u,v)=\frac{P_{iz-1/2}^{\mu }(u)P_{iz-1/2}^{-\mu
}(v)-P_{iz-1/2}^{-\mu }(u)P_{iz-1/2}^{\mu }(v)}{\sin (\mu \pi )},
\label{LegComb}
\end{equation}%
in the right-half plane of the complex variable $z$:%
\begin{equation}
X_{iz_{k}}^{\mu }(u,v)=0.  \label{Pzk0}
\end{equation}%
In (\ref{LegComb}), $P_{iz-1/2}^{\mu }(u)$\ is the associated Legendre
function\ of the first kind (in this paper the definition of the associated
Legendre functions follows that given in \cite{Abra72}). In the discussion
below we will assume that $u,v>1$. The expression in the numerator of (\ref%
{LegComb}) has simple zeros for integer values of $\mu $ and the function $%
X_{iz}^{\mu }(u,v)$ is regular at these points. Since one has the property $%
X_{\nu }^{-\mu }(u,v)=X_{\nu }^{\mu }(u,v)$, without loss of
generality, we consider the parameter $\mu $ being non-negative,
$\mu \geqslant 0$. For given values $u$, $v$, and $\mu $ the
function $X_{iz}^{\mu }(u,v)$ has an infinity of real zeros. From
the asymptotic formula for the associated
Legendre functions we can see that for $z\rightarrow +\infty $ one has%
\begin{equation}
X_{iz}^{\mu }(u,v)\approx \frac{2\sin [\left( \eta _{v}-\eta _{u}\right) z]}{%
\pi z\sqrt{\sinh \eta _{u}\sinh \eta _{v}}},  \label{Xlargez}
\end{equation}
where $\eta _{u}$ and $\eta _{v}$ are defined as%
\begin{equation}
u=\cosh \eta _{u},\;v=\cosh \eta _{v}.  \label{ueta}
\end{equation}%
From here we obtain the asymptotic expression for large zeros:%
\begin{equation}
z_{k}\approx \pi k/\left( \eta _{v}-\eta _{u}\right) .
\label{zkAsymp}
\end{equation}

In general, the zeros $z_{k}$ are functions of the parameters $u$, $v$, and $%
\mu $: $z_{k}=z_{k}(u,v,\mu )$. By taking into account that for the
associated Legendre function one has $P_{-\nu -1/2}^{\mu }(u)=P_{\nu
-1/2}^{\mu }(u)$, we see that $X_{-\nu }^{\mu }(u,v)=X_{\nu }^{\mu }(u,v)$.
Hence, the points $z=-z_{k}$ are zeros of the function $X_{iz}^{\mu }(u,v)$
as well. In Appendix \ref{sec:Zeros} we show that the zeros $z=z_{k}$ are
simple and under the conditions specified above the function $X_{iz}^{\mu
}(u,v)$ has no zeros which are not real. We will assume that $z_{k}$ are
arranged in ascending order of magnitude. Note that the function $X_{\nu
}^{\mu }(u,v)$ can also be expressed in terms of the combination
\begin{equation}
Y_{\nu }^{\mu }(u,v)=Q_{\nu -1/2}^{\mu }(u)P_{\nu -1/2}^{\mu }(v)-P_{\nu
-1/2}^{\mu }(u)Q_{\nu -1/2}^{\mu }(v),  \label{Ynu}
\end{equation}%
as%
\begin{equation}
X_{\nu }^{\mu }(u,v)=\frac{2}{\pi e^{i\mu \pi }}\frac{\Gamma ( \nu
-\mu +1/2) }{\Gamma ( \nu +\mu +1/2) }Y_{\nu }^{\mu }(u,v),
\label{XY}
\end{equation}%
where $Q_{\nu -1/2}^{\mu }(u)$ is the associated Legendre function of the
second kind and $\Gamma (x)$ is the gamma function.

A summation formula for the series over $z_{k}$ can be derived by using the
generalized Abel-Plana formula \cite{Sah1} (see also \cite{Saha00Rev,Saha07Rev}%
). For functions $f(z)$ and $g(z)$ meromorphic in the strip $a\leqslant
x\leqslant b$ of the complex plane $z=x+iy$ this formula has the form
\begin{eqnarray}
&&\lim_{b\rightarrow \infty }\bigg[{\mathrm{p.v.}}\!\int_{a}^{b}dx\,f(x)-\pi
i\sum_{k}\underset{z=z_{g,k}}{\mathrm{Res}}g(z)-\pi i\sum_{k,{{\mathrm{Im\,}}%
}z_{f,k}\neq 0}\sigma (z_{f,k})\underset{z={\mathrm{\,}}z_{f,k}}{\mathrm{Res}%
}f(z)\bigg]  \notag \\
&&\quad =\frac{1}{2}\int_{a-i\infty }^{a+i\infty }dz\,\left[ g(z)+{\sigma (z)%
}f(z)\right] ,  \label{GAPF}
\end{eqnarray}%
where ${\sigma (z)\equiv \mathrm{sgn}}({{\mathrm{Im\,}}}z)$ and p.v. means
the principal value of the integral. In this formula, $z_{f,k}$ and $z_{g,k}$
are the positions of the poles of the functions $f(z)$ and $g(z)$ in the
strip $a<x<b$. As functions $f(z)$ and $g(z)$ in formula (\ref{GAPF}) we
choose%
\begin{eqnarray}
f(z) &=&\frac{h(z)}{4Q_{iz-1/2}^{\mu }(u)Q_{-iz-1/2}^{\mu }(u)}\frac{\Gamma
\left( iz+\mu +1/2\right) \pi ^{2}e^{2i\mu \pi }i\sinh (z\pi )}{\Gamma
\left( iz-\mu +1/2\right) \cos [(iz-\mu )\pi ]},  \notag \\
g(z) &=&\left[ \frac{Q_{-iz-1/2}^{\mu }(v)}{Q_{-iz-1/2}^{\mu }(u)}+\frac{%
Q_{iz-1/2}^{\mu }(v)}{Q_{iz-1/2}^{\mu }(u)}\right] \frac{h(z)}{2X_{iz}^{\mu
}(u,v)},  \label{gz}
\end{eqnarray}%
where $h(z)$ is a meromorphic function for $a\leqslant {{\mathrm{Re}}}%
\,z\leqslant b$. The combinations appearing on the left-hand side of formula
(\ref{GAPF}) are presented in the form%
\begin{equation}
g(z)\pm f(z)=\frac{Q_{\mp iz-1/2}^{\mu }(v)}{Q_{\mp iz-1/2}^{\mu }(u)}\frac{%
h(z)}{X_{iz}^{\mu }(u,v)}.  \label{gzplmin}
\end{equation}%
Note that the function $g(z)$ has simple poles at the zeros $z_{k}$ of the
function (\ref{LegComb}).

The conditions for the generalized Abel-Plana formula (\ref{GAPF}),
formulated in terms of the function $h(z)$, take the form%
\begin{eqnarray}
\lim_{w\rightarrow \infty }\int_{a\pm iw}^{b\pm iw}dz\frac{Q_{\mp
iz-1/2}^{\mu }(v)}{Q_{\mp iz-1/2}^{\mu }(u)}\frac{h(z)}{X_{iz}^{\mu }(u,v)}
&=&0,  \notag \\
\lim_{b\rightarrow \infty }\int_{b}^{b\pm i\infty }dz\frac{Q_{\mp
iz-1/2}^{\mu }(v)}{Q_{\mp iz-1/2}^{\mu }(u)}\frac{h(z)}{X_{iz}^{\mu }(u,v)}
&=&0.  \label{Cond1}
\end{eqnarray}%
With the help of the asymptotic formulae for the associated Legendre
functions, we can see that these conditions are satisfied if the function $%
h(z)$ is restricted by the constraint%
\begin{equation}
|h(z)|<x^{-2\mu }\varepsilon (x)e^{c(\eta _{v}-\eta
_{u})y},\;z=x+iy,\;|z|\rightarrow \infty ,  \label{Cond2}
\end{equation}%
uniformly in any finite interval of \ $x$, where $c<2$, $\varepsilon
(x)\rightarrow 0$ for $x\rightarrow +\infty $.

Now, after the substitution of the functions (\ref{gz}) into formula (\ref%
{GAPF}), we see that for a function $h(z)$ meromorphic in the half-plane ${{%
\mathrm{Re}}}\,z\geqslant a$ and satisfying condition (\ref{Cond2}), the
following formula takes place%
\begin{eqnarray}
&&\lim_{b\rightarrow \infty }\bigg\{\sum_{k=m}^{n}\frac{h(z)}{\partial
_{z}X_{iz}^{\mu }(u,v)}\frac{Q_{iz-1/2}^{\mu }(v)}{Q_{iz-1/2}^{\mu }(u)}%
\bigg|_{z=z_{k}}+\frac{i}{\pi }{\mathrm{p.v.}}\!\int_{a}^{b}dx\,f(x)+r[h(z)]%
\bigg\}  \notag \\
&&\qquad =\frac{i}{2\pi }\int_{a-i\infty }^{a+i\infty }dz\,\frac{Q_{-\sigma
(z)iz-1/2}^{\mu }(v)}{Q_{-\sigma (z)iz-1/2}^{\mu }(u)}\frac{h(z)}{%
X_{iz}^{\mu }(u,v)},  \label{SumForm0}
\end{eqnarray}%
where the function $f(z)$ is defined by the relation (\ref{gz}). In this
formula we have introduced the notation%
\begin{eqnarray}
r[h(z)] &=&\sum_{k,{{\mathrm{Im\,}}}z_{h,k}\neq 0}\underset{z=z_{h,k}}{%
\mathrm{Res}}\bigg[\frac{Q_{-\sigma (z_{k})iz-1/2}^{\mu }(v)}{Q_{-\sigma
(z_{k})iz-1/2}^{\mu }(u)}\frac{h(z)}{X_{iz}^{\mu }(u,v)}\bigg]  \notag \\
&&+\frac{1}{2}\sum_{k,{{\mathrm{Im\,}}}z_{h,k}=0}\underset{z=z_{h,k}}{%
\mathrm{Res}}\bigg[\frac{h(z)}{X_{iz}^{\mu }(u,v)}\sum_{l=\pm }\frac{%
Q_{liz-1/2}^{\mu }(v)}{Q_{liz-1/2}^{\mu }(u)}\bigg].  \label{rhz}
\end{eqnarray}%
with $z_{h,k}$ being the positions of the poles for the function $h(z)$. On
the left-hand side of (\ref{SumForm0}), one has $z_{m-1}<a<z_{m}$, $%
z_{n}<b<z_{n+1}$ and in (\ref{rhz}) the summation goes over the poles $%
z_{h,k}$ in the strip $a<{{\mathrm{Re}}}\,z<b$. Note that one has the
relations%
\begin{equation}
\frac{Q_{iz-1/2}^{\mu }(v)}{Q_{iz-1/2}^{\mu }(u)}=\frac{P_{iz-1/2}^{\mu }(v)%
}{P_{iz-1/2}^{\mu }(u)}=\frac{P_{iz-1/2}^{-\mu }(v)}{P_{iz-1/2}^{-\mu }(u)}%
,\;z=z_{k},  \label{QWrons}
\end{equation}%
and in the summation of the first term in figure braces of (\ref{SumForm0})
we can replace the ratio of the associated Legendre functions of the second
kind by the ratio of the functions of the first kind.

A useful form of the summation formula (\ref{SumForm0}) is obtained in the
limit $a\rightarrow 0$. In this limit, we see that for a function $h(z)$
meromorphic in the half-plane ${{\mathrm{Re}}}\,z\geqslant 0$ and satisfying
the condition (\ref{Cond2}) the following formula holds%
\begin{eqnarray}
&&\sum_{k=1}^{\infty }\frac{h(z)}{\partial _{z}X_{iz}^{\mu }(u,v)}\frac{%
Q_{iz-1/2}^{\mu }(v)}{Q_{iz-1/2}^{\mu }(u)}\bigg|_{z=z_{k}}=\frac{\pi
e^{2i\mu \pi }}{4}{\mathrm{p.v.}}\!\int_{0}^{\infty }dx\,\frac{\Gamma \left(
ix+\mu +1/2\right) \sinh (x\pi )}{\Gamma \left( ix-\mu +1/2\right) \cos
[(ix-\mu )\pi ]}  \notag \\
&& \times \frac{h(x)}{Q_{ix-1/2}^{\mu }(u)Q_{-ix-1/2}^{\mu }(u)}%
-r[h(z)]-\frac{1}{2\pi }\int_{0}^{\infty }dx\,\frac{Q_{x-1/2}^{\mu }(v)}{%
Q_{x-1/2}^{\mu }(u)}\frac{h(xe^{\pi i/2})+h(xe^{-\pi i/2})}{X_{x}^{\mu }(u,v)%
}.  \label{SumFormula}
\end{eqnarray}%
For large values $x\gg 1$, for the associated Legendre functions we have
\begin{equation}
P_{x-1/2}^{\mu }(\cosh \eta )\approx \frac{x^{\mu -1/2}e^{\eta x}}{\sqrt{%
2\pi \sinh \eta }},\;Q_{x-1/2}^{\mu }(\cosh \eta )\approx \sqrt{\frac{\pi }{2%
}}e^{i\mu \pi }\frac{x^{\mu -1/2}e^{-\eta x}}{\sqrt{\sinh \eta }}.
\label{PQasimp}
\end{equation}%
By using these formulae and the relation (\ref{XY}), for the corresponding
asymptotic behavior of the function $X_{x}^{\mu }(u,v)$ one finds
\begin{equation}
X_{x}^{\mu }(u,v)\approx \frac{e^{(\eta _{v}-\eta _{u})x}}{\pi x\sqrt{\sinh
\eta _{u}\sinh \eta _{v}}}.  \label{Xxlarge}
\end{equation}%
From these asymptotic formulae it follows that under the condition (\ref%
{Cond2}) for the function $h(z)$, the second integral on the right-hand side
of formula (\ref{SumFormula}) exponentially converges in the upper limit.

If the function $h(z)$ has poles on the positive real axis, it is assumed
that the first integral on the right-hand side converges in the sense of the
principal value. From the derivation of (\ref{SumFormula}) it follows that
this formula may be extended to the case of some functions $h(z)$ having
branch-points on the imaginary axis, for example, having the form $%
h(z)=h_{1}(z)/(z^{2}+c^{2})^{1/2}$, where $h_{1}(z)$ is a
meromorphic function. This type of function appears in the
physical example discussed in section \ref{sec:Phys}. Special
cases of formula (\ref{SumFormula}) are considered in the next
section.

Another generalization of formula (\ref{SumFormula}) can be given for a
class of functions $h(z)$ having purely imaginary poles at the points $z=\pm
iy_{k}$, $y_{k}>0$, $k=1,2,\ldots $, and at the origin $z=y_{0}=0$. We
assume that the function $h(z)$ satisfies the condition%
\begin{equation}
h(z)=-h(ze^{-\pi i})+o((z-\sigma _{k})^{-1}),\;z\rightarrow \sigma
_{k},\;\sigma _{k}=0,iy_{k}.  \label{Impolecond}
\end{equation}%
Let us denote by $C_{\rho }(\sigma _{k})$ the right half of the circle with
radius $\rho $ and with the center at the point $\sigma _{k}$, described in
the positive direction. Similarly, we denote by $\gamma _{\rho }^{+}$ and $%
\gamma _{\rho }^{-}$ the upper and lower halves of the semicircle in the
right half-plane with radius $\rho $ and with the center at the point $z=0$,
described in the positive direction with respect to this point. Now, in the
limit $a\rightarrow 0$ the right-hand side of (\ref{SumForm0}) can be
presented in the form%
\begin{equation}
\frac{i}{2\pi }\sum_{\alpha =+,-}\bigg(\int_{\gamma _{\rho }^{\alpha
}}dz+\sum_{\sigma _{k}=\alpha iy_{k}}\int_{C_{\rho }(\sigma _{k})}dz\bigg)\,%
\frac{Q_{-\alpha iz-1/2}^{\mu }(v)}{Q_{-\alpha iz-1/2}^{\mu }(u)}\frac{h(z)}{%
X_{iz}^{\mu }(u,v)},  \label{Impoles1}
\end{equation}%
plus the sum of the integrals along the straight segments $(\pm
i(y_{k-1}+\rho ),\pm i(y_{k}-\rho ))$\ of the imaginary axis between the
poles. In the limit $\rho \rightarrow 0$ the sum of the integrals along the
straight segments of the imaginary axis gives the principal value of the
last integral on the right-hand side of (\ref{SumFormula}). In the terms of (%
\ref{Impoles1}) with $\alpha =-$ we introduce a new integration variable $%
z^{\prime }=ze^{\pi i}$. By using the relation (\ref{Impolecond}), the
expression (\ref{Impoles1}) is presented in the form%
\begin{equation}
-\sum_{\sigma _{k}=0,iy_{k}}(1-\delta _{0\sigma _{k}}/2)\underset{z=\sigma
_{k}}{\mathrm{Res}}\bigg[\frac{Q_{-iz-1/2}^{\mu }(v)}{Q_{-iz-1/2}^{\mu }(u)}%
\frac{h(z)}{X_{iz}^{\mu }(u,v)}\bigg]  \label{Impoles2}
\end{equation}%
plus the part which vanishes in the limit $\rho \rightarrow 0$. As a result,
formula (\ref{SumFormula}) is extended for functions having purely imaginary
poles and satisfying condition (\ref{Impolecond}). For this, on the
right-hand side of (\ref{SumFormula}) we have to add the sum of residues (%
\ref{Impoles2}) at these poles and take the principal value of the second
integral on the right-hand side. The latter exists due to condition (\ref%
{Impolecond}).

\section{Special cases}

\label{sec:Special}

First we consider \ the case $\mu =1/2$. For the corresponding associated
Legendre functions one has%
\begin{equation}
P_{z-1/2}^{-1/2}(\cosh \eta )=\sqrt{\frac{2}{\pi }}\frac{\sinh (z\eta )}{z%
\sqrt{\sinh \eta }},\;P_{z-1/2}^{1/2}(\cosh \eta )=\sqrt{\frac{2}{\pi }}%
\frac{\cosh (z\eta )}{\sqrt{\sinh \eta }}.  \label{SpCase1}
\end{equation}%
By making use of these formulae we find
\begin{equation}
X_{iz}^{1/2}(u,v)=\frac{2}{\pi }\frac{\sin [z(\eta _{v}-\eta _{u})]}{z\sqrt{%
\sinh \eta _{u}\sinh \eta _{v}}}.  \label{X1/2}
\end{equation}%
Hence, in this case for the zeros $z_{k}$ one has $z_{k}=\pi k/(\eta
_{v}-\eta _{u})$. Introducing a new function $F(z)$ in accordance with the
relation $zh(z)=F(z(\eta _{v}-\eta _{u})/\pi )$, from the formula (\ref%
{SumFormula}) we obtain the Abel-Plana summation formula in its standard
form:%
\begin{equation}
\sum_{k=1}^{\infty }F(k)=-\frac{1}{2}F(0)+\int_{0}^{\infty
}dx\,F(x)+i\int_{0}^{\infty }dx\frac{F(ix)-F(-ix)}{e^{2\pi x}-1},
\label{AP1}
\end{equation}%
where the first term on the right-hand side comes from the residue term at $%
\sigma _{k}=0$ in (\ref{Impoles2}).

Now let us show that from formula (\ref{SumFormula}), as a special case, a
summation formula is obtained for the series over zeros of the combination
of cylinder functions. First of all, by making use of formulae
\begin{equation}
\lim_{s\rightarrow +\infty }(sz)^{\pm \mu }P_{isz-1/2}^{\mp \mu }(\cosh
(\lambda /s))=J_{\pm \mu }(\lambda z),  \label{Plim}
\end{equation}%
with $J_{\mu }(\eta )$ being the Bessel function of the first kind, we can
see that the following relation holds:%
\begin{equation}
\lim_{s\rightarrow +\infty }X_{isz}^{\mu }(\cosh (\lambda _{u}/s),\cosh
(\lambda _{v}/s))=C_{\mu }(\lambda _{u}z,\lambda _{v}z),  \label{XBess}
\end{equation}%
where%
\begin{equation}
C_{\mu }(\lambda _{u}z,\lambda _{v}z)=J_{\mu }(\lambda _{u}z)Y_{\mu
}(\lambda _{v}z)-Y_{\mu }(\lambda _{u}z)J_{\mu }(\lambda _{v}z).  \label{Cmu}
\end{equation}%
Note that, instead of the function $J_{-\mu }(z)$ we have
introduced the Neumann function $Y_{\mu }(z)$. Hence, in the limit
$s\rightarrow \infty $ from (\ref{SumFormula}) we obtain the
summation formula for the series over zeros $z=\lambda _{\mu ,k}$,
$k=1,2,\ldots $, of the function $C_{\mu
}(\lambda _{u}z,\lambda _{v}z)$. For this, first we rewrite formula (\ref%
{SumFormula}) making the replacements $z\rightarrow sz$, $x\rightarrow sx$,
in both sides of this formula including the terms in $r[h(z)]$, and we take $%
u=\cosh (\lambda _{u}/s)$, $v=\cosh (\lambda _{v}/s)$. Introducing a new
function $F(z)=h(sz)$, in the limit $s\rightarrow +\infty $ we find the
formula
\begin{eqnarray}
&&\sum_{k=1}^{\infty }\frac{F(z)}{\partial _{z}C_{\mu }(\lambda
_{u}z,\lambda _{v}z)}\frac{J_{\mu }(\lambda _{v}z)}{J_{\mu }(\lambda _{u}z)}%
\bigg|_{z=\lambda _{\mu ,k}}=\frac{1}{\pi
}{\mathrm{p.v.}}\!\int_{0}^{\infty }dx\,\frac{F(x)}{J_{\mu
}^{2}(\lambda _{u}x)+Y_{\mu }^{2}(\lambda _{u}x)}
\notag \\
&&\qquad -r_{C}[F(z)]-\frac{1}{4}\int_{0}^{\infty }dx\,\frac{K_{\mu
}(\lambda _{v}x)}{K_{\mu }(\lambda _{u}x)}\frac{F(xe^{\pi i/2})+F(xe^{-\pi
i/2})}{K_{\mu }(\lambda _{u}x)I_{\mu }(\lambda _{v}x)-I_{\mu }(\lambda
_{u}x)K_{\mu }(\lambda _{v}x)},  \label{SumFormBess}
\end{eqnarray}%
where $I_{\mu }(x)$ and $K_{\mu }(x)$\ are the modified Bessel functions and%
\begin{eqnarray}
r_{C}[F(z)] &=&\pi \sum_{k}\underset{{{\mathrm{Im\,}}}z_{F,k}=0}{\mathrm{Res}%
}\bigg[\frac{J_{\mu }(\lambda _{u}z)J_{\mu }(\lambda _{v}z)+Y_{\mu }(\lambda
_{u}z)Y_{\mu }(\lambda _{v}z)}{J_{\mu }^{2}(\lambda _{u}x)+Y_{\mu
}^{2}(\lambda _{u}x)}\frac{F(z)}{C_{\mu }(\lambda _{u}z,\lambda _{v}z)}\bigg]
\notag \\
&&+\pi \sum_{l=1,2}\sum_{k}\underset{(-1)^{l}{{\mathrm{Im\,}}}z_{F,k}<0}{%
\mathrm{Res}}\bigg[\frac{H_{\mu }^{(l)}(\lambda _{v}z)}{H_{\mu
}^{(l)}(\lambda _{u}z)}\frac{F(z)}{C_{\mu }(\lambda _{u}z,\lambda _{v}z)}%
\bigg].  \label{rCF}
\end{eqnarray}%
In deriving (\ref{SumFormBess}) we have also used the formulae
\begin{eqnarray}
\lim_{\nu \rightarrow +\infty }\nu ^{-\mu }Q_{i\nu -1/2}^{\mu }(\cosh (\eta
/\nu )) &=&-\frac{\pi i}{2}e^{i\mu \pi }H_{\mu }^{(2)}(\eta ),  \notag \\
\lim_{\nu \rightarrow \infty }\nu ^{\pm \mu }P_{\nu }^{\mp \mu
}(\cosh
(x/\nu )) &=&I_{\pm \mu }(x),  \label{LegLimit} \\
\lim_{\nu \rightarrow \infty }\nu ^{-\mu }Q_{\nu }^{\mu }(\cosh
(x/\nu )) &=&e^{i\mu \pi }K_{\mu }(x),  \notag
\end{eqnarray}%
and the relation
\begin{equation}
\frac{H_{\mu }^{(2)}(\lambda _{v}z)}{H_{\mu }^{(2)}(\lambda _{u}z)}=\frac{%
J_{\mu }(\lambda _{v}z)}{J_{\mu }(\lambda _{u}z)},\;z=\lambda
_{\mu ,k}.
\end{equation}%
Note that from (\ref{LegLimit}) it follows that
\begin{equation}
\lim_{s\rightarrow +\infty }X_{sx}^{\mu }(\cosh (\lambda _{u}/s),\cosh
(\lambda _{v}/s))=\frac{2}{\pi }\left[ K_{\mu }(\lambda _{u}x)I_{\mu
}(\lambda _{v}x)-I_{\mu }(\lambda _{u}x)K_{\mu }(\lambda _{v}x)\right] .
\label{rel5}
\end{equation}%
Formula (\ref{SumFormBess}) is a special case of the result derived in \cite%
{Sah1} (see also, \cite{Saha07Rev}). Physical applications of this
formula are given in \cite{Saha01,Saha04}.

\section{Vacuum polarization by concentric spherical boundaries in a constant
curvature space}

\label{sec:Phys}

In this section we give a physical application of the summation formula (\ref%
{SumFormula}). Consider a scalar field $\varphi (x)$ on background of the
space with constant negative curvature described by the line element
\begin{equation}
ds^{2}=dt^{2}-a^{2}\left[ dr^{2}+\sinh ^{2}r(d\theta ^{2}+\sin ^{2}\theta
d\phi ^{2})\right] ,  \label{metric}
\end{equation}%
where $a$ is a constant. The field equation has the form%
\begin{equation}
\left( \nabla _{l}\nabla ^{l}+M^{2}+\xi R\right) \varphi (x)=0,
\label{FieldEq}
\end{equation}%
where $M$ is the mass of the field quanta, $\xi $ is the curvature coupling
parameter and for the Ricci scalar one has $R=-6a^{-2}$. We will assume that
the field operator satisfies Dirichlet boundary conditions on two concentric
spherical shells with radii $r=r_{1}$ and $r=r_{2}$, $r_{1}<r_{2}$,
\begin{equation}
\varphi (x)|_{r=r_{1,2}}=0.  \label{BoundCond}
\end{equation}

The boundary conditions modify the spectrum of the zero-point
fluctuations and, as a result of this modification, the physical
properties of the vacuum are changed. Among the most important
characteristics of these properties are the expectation values of
quantities bilinear in the field operator such as the field
squared and the energy-momentum tensor. These expectation values
are obtained from two-point functions in the coincidence limit of
the arguments. As a two-point function here we will consider the
positive frequency Wightman function. Other two-point functions
are evaluated in a similar way. Expanding the field operator over
the complete set $\{\varphi _{\alpha }(x),\varphi _{\alpha }^{\ast
}(x)\}$ of classical solutions to the field equation satisfying
the boundary conditions (\ref{BoundCond}), the
Wightman function is presented in the form of the following mode-sum%
\begin{equation}
W(x,x^{\prime })=\langle 0|\varphi (x)\varphi (x^{\prime })|0\rangle
=\sum_{\alpha }\varphi _{\alpha }(x)\varphi _{\alpha }^{\ast }(x^{\prime }),
\label{WFsum}
\end{equation}%
where $|0\rangle $ is the amplitude of the vacuum state and $\alpha $ is a
set of quantum numbers specifying the solution.

In accordance with the spherical symmetry of the problem under
consideration, the eigenfunctions for the scalar field can be
presented in the factorized form
\begin{equation}
\varphi _{\alpha }(x)=Z(r)Y_{lm}(\theta ,\phi )e^{-i\omega t},
\label{eigfunc1}
\end{equation}%
where $Y_{lm}(\theta ,\phi )$ are the spherical harmonics with $%
l=0,1,2,\ldots $, $-l\leqslant m\leqslant l$. The equation for the radial
function is obtained from the field equation (\ref{FieldEq}) and has the form%
\begin{equation}
\frac{1}{\sinh ^{2}r}\frac{d}{dr}\left( \sinh ^{2}r\frac{dZ}{dr}\right) +%
\left[ (\omega ^{2}-M^{2})a^{2}+6\xi -\frac{l(l+1)}{\sinh ^{2}r}\right] Z=0.
\label{Zeq}
\end{equation}%
In the region between the spherical shells the solution of equation (\ref%
{Zeq}) is expressed in terms of the associated Legendre function as%
\begin{equation*}
Z(r)=\frac{c_{1}P_{iz-1/2}^{-l-1/2}(u)+c_{2}P_{iz-1/2}^{l+1/2}(u)}{\sqrt{\sinh r%
}},
\end{equation*}%
with integration constants $c_{1}$ and $c_{2}$ and the notations%
\begin{equation}
z^{2}=(\omega ^{2}-M^{2})a^{2}+6\xi -1,\;u=\cosh r.  \label{lambda}
\end{equation}

From the boundary condition on the inner sphere we find%
\begin{equation}
\frac{c_{2}}{c_{1}}=-\frac{P_{iz-1/2}^{-l-1/2}(u_{1})}{%
P_{iz-1/2}^{l+1/2}(u_{1})},\;u_{i}\equiv \cosh r_{i},\;i=1,2,
\label{Pu1Ratio}
\end{equation}%
and, hence,%
\begin{equation}
Z(r)=C_{\alpha }\frac{X_{iz}^{l+1/2}(u_{1},u)}{\sqrt{\sinh r}},
\label{ZX}
\end{equation}%
where $C_{\alpha }$ is the normalization constant and the function $%
X_{iz}^{l+1/2}(u_{1},u)$ is defined by (\ref{LegComb}). From the boundary
condition on the outer sphere we see that the eigenvalues for $z$ are
solutions of the equation%
\begin{equation}
X_{iz}^{l+1/2}(u_{1},u_{2})=0.  \label{ejgmodes}
\end{equation}

As a result, the eigenfunctions have the form%
\begin{equation}
\varphi _{\alpha }(x)=\frac{C_{\alpha }}{\sqrt{\sinh r}}%
X_{iz}^{l+1/2}(u_{1},u)Y_{lm}(\theta ,\phi )e^{-i\omega t},  \label{eigfunc2}
\end{equation}%
and, hence, $z=z_{k}$, $k=1,2,\ldots $, in the notations of section \ref%
{sec:SumForm}. The corresponding eigenfrequencies are related to these zeros
by the formula%
\begin{equation}
\omega _{k}^{2}=\omega ^{2}(z_{k})=(z_{k}^{2}+1-6\xi )/a^{2}+M^{2}.
\label{eigfreq}
\end{equation}%
Hence, the set $\alpha $ of the quantum numbers is specified to $\alpha
=(l,m,k)$.

The coefficient $C_{\alpha }$ in (\ref{eigfunc2}) is determined from the
orthonormalization condition for the eigenfunctions:%
\begin{equation}
\int d^{3}x\,\sqrt{|g|}\varphi _{\alpha }(x)\varphi _{\alpha ^{\prime
}}^{\ast }(x)=\frac{\delta _{\alpha \alpha ^{\prime }}}{2\omega },
\label{normcond}
\end{equation}%
where the integration goes over the region between the spherical shells.
Making use of the integration formula given in Appendix \ref{sec:Zeros} and
the boundary conditions, for this coefficient we find%
\begin{equation}
C_{\alpha }^{-2}=a^{3}\frac{\omega (z)}{z}(u_{2}^{2}-1)[\partial
_{z}X_{iz}^{l+1/2}(u_{1},u_{2})]\partial _{u}X_{iz}^{l+1/2}(u_{1},u),
\label{Cnorm}
\end{equation}%
with $z=z_{k}$, $u=u_{2}$. By using the Wronskian relation for the
associated Legendre functions,%
\begin{equation}
W\{P_{i\nu -1/2}^{\mu }(u),Q_{i\nu -1/2}^{\mu }(u)\}=\frac{e^{i\mu \pi
}\Gamma (i\nu +\mu +1/2)}{(1-u^{2})\Gamma (i\nu -\mu +1/2)},  \label{WrPQ}
\end{equation}%
it can be seen that
\begin{equation}
\lbrack \partial _{u}X_{iz_{k}}^{l+1/2}(u_{1},u)]_{u=u_{2}}=\frac{2}{\pi }%
\frac{1}{u_{2}^{2}-1}\frac{P_{iz_{k}-1/2}^{l+1/2}(u_{1})}{%
P_{iz_{k}-1/2}^{l+1/2}(u_{2})}.  \label{dXrel1}
\end{equation}%
Upon the substitution this into (\ref{Cnorm}), the normalization coefficient
is written in the equivalent form%
\begin{equation}
C_{\alpha }^{-2}=a^{3}\frac{2\omega (z)}{\pi z}\partial
_{z}X_{iz}^{l+1/2}(u_{1},u_{2})\frac{P_{iz-1/2}^{l+1/2}(u_{1})}{%
P_{iz-1/2}^{l+1/2}(u_{2})}\bigg|_{z=z_{k}}.  \label{Cnorm2}
\end{equation}%
Note that the ratio of the gamma functions in this formula can also be
presented in the form%
\begin{equation}
\frac{\Gamma (iz_{k}+l+1)}{\Gamma (iz_{k}-l)}=\frac{1}{\pi }\cos
[\pi (iz_{k}-l-1/2)]|\Gamma (iz_{k}+l+1)|^{2}.  \label{GamRatio}
\end{equation}

Substituting the eigenfunctions into the mode-sum formula (\ref{WFsum}) and
using the addition theorem for the spherical harmonics, for the Wightman
function one finds%
\begin{eqnarray}
W(x,x^{\prime }) &=&\frac{1}{8a^{3}}\sum_{l=0}^{\infty }\frac{%
(2l+1)P_{l}(\cos \gamma )}{\sqrt{\sinh r\sinh r^{\prime }}}  \notag \\
&&\times \sum_{k=1}^{\infty }z\frac{%
X_{iz}^{l+1/2}(u_{1},u)X_{iz}^{l+1/2}(u_{1},u^{\prime })}{\partial
_{z}X_{iz}^{l+1/2}(u_{1},u_{2})}\frac{P_{iz-1/2}^{l+1/2}(u_{2})}{%
P_{iz-1/2}^{l+1/2}(u_{1})}\frac{e^{-i\omega (z)\Delta t}}{\omega (z)}\bigg|%
_{z=z_{k}}  \label{WF1}
\end{eqnarray}%
where $\Delta t=t-t^{\prime }$ and $u^{\prime }=\cosh r^{\prime }$. In (\ref%
{WF1}), $P_{l}(\cos \gamma )$ is the Legendre polynomial and
\begin{equation}
\cos \gamma =\cos \theta \cos \theta ^{\prime }+\sin \theta \sin \theta
^{\prime }\cos (\phi -\phi ^{\prime }).  \label{cosgam}
\end{equation}%
As the expressions for the zeros $z_{k}$ are not explicitly known, formula (%
\ref{WF1}) for the Wightman function is not convenient. In addition, the
terms in the sum are highly oscillatory for large values of quantum numbers.

For the further evaluation of the Wightman function we apply to the series
over $k$ the summation formula (\ref{SumFormula}) with $u=u_{1}$ and $%
v=u_{2} $, taking in this formula%
\begin{equation}
h(z)=zX_{iz}^{l+1/2}(u_{1},u)X_{iz}^{l+1/2}(u_{1},u^{\prime })\frac{%
e^{-i\omega (z)\Delta t}}{\omega (z)},  \label{hz}
\end{equation}%
where the function $\omega (z)$ is defined by (\ref{eigfreq}). The function (%
\ref{hz}) has no poles in the right-half plane and, hence, $r[h(z)]=0$. The
corresponding conditions are satisfied if $r+r^{\prime }+\Delta t/a<2r_{2}$.
In particular, this is the case in the coincidence limit $t=t^{\prime }$ for
the region under consideration. For the function (\ref{hz}) the part of the
integral on the right-hand side of formula (\ref{SumFormula}) over the
region $(0,x_{M})$ vanishes, and for the Wightman function one finds%
\begin{eqnarray}
W(x,x^{\prime }) &=&W_{1}(x,x^{\prime })-\frac{1}{8\pi a^{2}}%
\sum_{l=0}^{\infty }\frac{(2l+1)P_{l}(\cos \gamma )}{\sqrt{\sinh r\sinh
r^{\prime }}}\int_{x_{M}}^{\infty }dx\,x  \notag \\
&&\times \frac{Q_{x-1/2}^{l+1/2}(u_{2})}{Q_{x-1/2}^{l+1/2}(u_{1})}\frac{%
X_{x}^{l+1/2}(u_{1},u)X_{x}^{l+1/2}(u_{1},u^{\prime })}{%
X_{x}^{l+1/2}(u_{1},u_{2})}\frac{\cosh (\sqrt{x^{2}-x_{M}^{2}}\Delta t/a)}{%
\sqrt{x^{2}-x_{M}^{2}}},  \label{WF2}
\end{eqnarray}%
where we have defined%
\begin{equation}
x_{M}=\sqrt{M^{2}a^{2}+1-6\xi }.  \label{xM}
\end{equation}%
In formula (\ref{WF2}), the first term on the right-hand side is given by%
\begin{eqnarray}
W_{1}(x,x^{\prime }) &=&-\frac{1}{32a^{3}}\sum_{l=0}^{\infty }\frac{%
(2l+1)P_{l}(\cos \gamma )}{\sqrt{\sinh r\sinh r^{\prime }}}\int_{0}^{\infty
}dx\,x\sinh (x\pi )  \notag \\
&&\times |\Gamma (ix+l+1)|^{2}\frac{%
X_{ix}^{l+1/2}(u_{1},u)X_{ix}^{l+1/2}(u_{1},u^{\prime })}{%
Q_{ix-1/2}^{l+1/2}(u_{1})Q_{-ix-1/2}^{l+1/2}(u_{1})}\frac{e^{-i\omega
(x)\Delta t}}{\omega (x)}.  \label{WF0}
\end{eqnarray}%
This function does not depend on the outer sphere radius whereas the second
term in (\ref{WF2}) vanishes in the limit $r_{2}\rightarrow \infty $. Hence,
the two-point function given by (\ref{WF0}) is the Wightman function for a
scalar field in background spacetime described by the line element (\ref%
{metric}) outside a single sphere with radius $r_{1}$ on which the
field obeys Dirichlet boundary condition. This can also be seen by
the direct evaluation using the corresponding eigenfunctions.
Thus, we can interpret the second term on the right-hand side of
(\ref{WF2}) as the part in the Wightman function induced by the
presence of the outer sphere.

An alternative form for the function (\ref{WF0}) is obtained by making use
of the identity%
\begin{eqnarray}
\frac{X_{ix}^{l+1/2}(u_{1},u)X_{ix}^{l+1/2}(u_{1},u^{\prime })}{%
Q_{ix-1/2}^{l+1/2}(u_{1})Q_{-ix-1/2}^{l+1/2}(u_{1})} &=&-\frac{4}{\pi ^{2}}%
P_{ix-1/2}^{-l-1/2}(u)P_{ix-1/2}^{-l-1/2}(u^{\prime })  \notag \\
&&-\frac{4i}{\pi ^{3}}P_{ix-1/2}^{-l-1/2}(u_{1})\sum_{\sigma =\pm 1}\frac{%
Q_{\sigma ix-1/2}^{-l-1/2}(u)Q_{\sigma ix-1/2}^{-l-1/2}(u^{\prime })}{%
Q_{\sigma ix-1/2}^{-l-1/2}(u_{1})}.  \label{Ident1}
\end{eqnarray}%
Substituting (\ref{Ident1}) into (\ref{WF0}), we can see that the part with
the first term on the right of formula~(\ref{Ident1}),%
\begin{eqnarray}
W_{0}(x,x^{\prime }) &=&\frac{1}{8\pi ^{2}a^{3}}\sum_{l=0}^{\infty }\frac{%
(2l+1)P_{l}(\cos \gamma )}{\sqrt{\sinh r\sinh r^{\prime }}}\int_{0}^{\infty
}dx\,x\sinh (\pi x)  \notag \\
&&\times |\Gamma (ix+l+1)|^{2}P_{ix-1/2}^{-l-1/2}(\cosh
r)P_{ix-1/2}^{-l-1/2}(\cosh r^{\prime })\frac{e^{-i\omega (x)\Delta t}}{%
\omega (x)},  \label{WF00}
\end{eqnarray}%
is the Wightman function for a scalar field on background of the constant
curvature space without boundaries (see \cite{Saha08}). In the part with the
second term on the right-hand side of formula (\ref{Ident1}) we rotate the
contour of integration over $x$ by the angle $\pi /2$ for the term with $%
\sigma =-1$ and by the angle $-\pi /2$ \ for the term with $\sigma =1$. As a
result, the exterior Wightman function for a single spherical boundary is
presented in the decomposed form
\begin{eqnarray}
W_{1}(x,x^{\prime }) &=&W_{0}(x,x^{\prime })-\frac{i}{4\pi ^{2}a^{2}}%
\sum_{l=0}^{\infty }(-1)^{l}\frac{(2l+1)P_{l}(\cos \gamma )}{\sqrt{\sinh
r\sinh r^{\prime }}}\int_{x_{M}}^{\infty }dx\,x\frac{\Gamma (x+l+1)}{\Gamma (x-l)%
}  \notag \\
&&\times \frac{P_{x-1/2}^{-l-1/2}(u_{1})}{Q_{x-1/2}^{-l-1/2}(u_{1})}%
Q_{x-1/2}^{-l-1/2}(\cosh r)Q_{x-1/2}^{-l-1/2}(\cosh r^{\prime })\frac{\cosh (%
\sqrt{x^{2}-x_{M}^{2}}\Delta t/a)}{\sqrt{x^{2}-x_{M}^{2}}},  \label{WF1ext}
\end{eqnarray}%
where the second term on the right-hand side is induced by the spherical
boundary. The Wightman function for the region inside a single spherical
shell is investigated in \cite{Saha08}. The corresponding expression is
obtained from (\ref{WF1ext}) by the replacements $P_{x-1/2}^{-l-1/2}%
\rightleftarrows Q_{x-1/2}^{-l-1/2}$ in the second term on the right of this
formula.

Taking the limit $a\rightarrow \infty $ with fixed $ar=R$, from the formulae
given above we obtain the corresponding results for spherical boundaries in
the Minkowski spacetime with radii $R_{1}=ar_{1}$ and $R_{2}=ar_{2}$. Note
that in this limit one has $x_{M}=aM$ and the result does not depend on the
curvature coupling parameter. Introducing a new integration variable $y=x/a$
and using the asymptotic formula for the gamma function for large values of
the argument, from (\ref{WF2}) we find%
\begin{eqnarray}
W^{\mathrm{(M)}}(x,x^{\prime }) &=&W_{1}^{\mathrm{(M)}}(x,x^{\prime
})-\sum_{l=0}^{\infty }\frac{(2l+1)P_{l}(\cos \gamma )}{4\pi ^{2}\sqrt{%
RR^{\prime }}}\int_{M}^{\infty }dy\,y\frac{\cosh (\sqrt{y^{2}-M^{2}}\Delta t)%
}{\sqrt{y^{2}-M^{2}}}  \notag \\
&&\times \frac{K_{l+1/2}(R_{2}y)}{K_{l+1/2}(R_{1}y)}\frac{%
G_{l+1/2}(R_{1}y,Ry)G_{l+1/2}(R_{1}y,R^{\prime }y)}{G_{l+1/2}(R_{1}y,R_{2}y)}%
,  \label{WFM}
\end{eqnarray}%
where we have introduced the notation $G_{\nu }(x,y)=K_{\nu }(x)I_{\nu
}(y)-K_{\nu }(y)I_{\nu }(x)$. The first term on the right-hand side of
formula (\ref{WFM}) is the Wightman function in the region outside a single
spherical boundary with radius $R_{1}$ in the Minkowski bulk. This function
is given by the expression%
\begin{eqnarray}
W_{1}^{\mathrm{(M)}}(x,x^{\prime }) &=&W_{0}^{\mathrm{(M)}}(x,x^{\prime
})-\sum_{l=0}^{\infty }\frac{(2l+1)P_{l}(\cos \gamma )}{4\pi ^{2}\sqrt{%
RR^{\prime }}}\int_{M}^{\infty }dy\,y  \notag \\
&&\times K_{l+1/2}(Ry)K_{l+1/2}(R^{\prime }y)\frac{I_{l+1/2}(R_{1}y)}{%
K_{l+1/2}(R_{1}y)}\frac{\cosh (\sqrt{y^{2}-M^{2}}\Delta t)}{\sqrt{y^{2}-M^{2}%
}}.  \label{WMink}
\end{eqnarray}%
Expressions (\ref{WFM}) and (\ref{WMink}) are special cases of the general
formulae given in \cite{Saha01} for a scalar field with Robin boundary
conditions in arbitrary number of spatial dimensions.

The vacuum expectation value of the field squared is obtained from
the Wightman function taking the coincidence limit of the
arguments. This limit is divergent and some renormalization
procedure is necessary. Here the important point is that for
points outside the spherical shells the local geometry is the same
as for the case of without boundaries and, hence, the structure of
the divergences is the same as well. This is also directly seen
from formulae (\ref{WF2}) and (\ref{WF1ext}), where the second
terms on the right-hand sides are finite in the coincidence limit.
Since in these formulae we have already explicitly subtracted the
boundary-free part, the renormalization is reduced to that for the
geometry without boundaries. In this way for the renormalized
vacuum expectation value of the field squared
one has%
\begin{eqnarray}
\langle \varphi ^{2}\rangle _{\mathrm{ren}} &=&\langle \varphi ^{2}\rangle
_{1,\mathrm{ren}}+\frac{\pi }{16a^{2}}\sum_{l=0}^{\infty }\frac{2l+1}{\sinh r%
}\int_{x_{M}}^{\infty }dx\,\frac{x}{\sqrt{x^{2}-x_{M}^{2}}}  \notag \\
&&\times \frac{\Gamma (x+l+1)}{\Gamma (x-l)}\frac{Q_{x-1/2}^{l+1/2}(u_{2})}{%
Q_{x-1/2}^{l+1/2}(u_{1})}\frac{[X_{x}^{l+1/2}(u_{1},u)]^{2}}{%
X_{x}^{l+1/2}(u_{1},u_{2})},  \label{phi2}
\end{eqnarray}%
where the first term on the right-hand side is the corresponding quantity
outside a spherical boundary with radius $r_{1}$ in the constant negative
curvature space without boundaries and the second one is induced by the
presence of the second spherical shell with the radius $r_{2}$. For the
first term one has%
\begin{eqnarray}
\langle \varphi ^{2}\rangle _{1,\mathrm{ren}} &=&\langle \varphi ^{2}\rangle
_{0,\mathrm{ren}}-\sum_{l=0}^{\infty }\frac{e^{i(l+1/2)\pi }}{4\pi ^{2}a^{2}}%
\frac{(2l+1)}{\sinh r}\int_{x_{M}}^{\infty }dx\,x  \notag \\
&&\times \frac{\Gamma (x+l+1)}{\Gamma (x-l)}\frac{P_{x-1/2}^{-l-1/2}(u_{1})}{%
Q_{x-1/2}^{-l-1/2}(u_{1})}\frac{\left[ Q_{x-1/2}^{-l-1/2}(u)\right] ^{2}}{%
\sqrt{x^{2}-x_{M}^{2}}},  \label{phi2single}
\end{eqnarray}%
where $\langle \varphi ^{2}\rangle _{0,\mathrm{ren}}$ is the vacuum
expectation value\ for the field squared in the constant negative curvature
space without boundaries and the second one is induced by the presence of a
single spherical shell with radius $r_{1}$. Note that the corresponding
formula for the vacuum expectation value inside a spherical shell (see \cite%
{Saha08}) is obtained from (\ref{phi2single}) by the replacements $%
P_{x-1/2}^{-l-1/2}\rightleftarrows Q_{x-1/2}^{-l-1/2}$ in the
second term on the right-hand side.

The physical example discussed in this section demonstrates the
advantages for the application of the Abel-Plana-type formulae in
the evaluation of the expectation values of local physical
observables in the presence of boundaries. For the summation of
the corresponding mode-sums the explicit form of the
eigenfrequencies is not necessary and the part corresponding to
the boundary-free space is explicitly extracted. Further, the
boundary induced parts are presented in the form of integrals
which rapidly converge and are finite in the coincidence limit for
points away from the boundaries. In this way the renormalization
procedure for local physical observables is reduced to that in
quantum field theory without boundaries. Methods for the
evaluation of global characteristics of the vacuum, such as the
total Casimir energy, in problems where the eigenmodes are given
implicitly as zeros of a given function, are described in
references~\cite{Eliz94}.

\section{Conclusion}

\label{sec:Conclus}

The associated Legendre functions arise in many problems of
mathematical physics. By making use of the generalized Abel-Plana
formula, we have derived summation formula (\ref{SumFormula}) for
the series over the zeros of the combination (\ref{LegComb}) of
the associated Legendre functions with respect to the degree. This
formula is valid for functions $h(z)$ meromorphic in the right
half-plane and obeying condition (\ref{Cond2}). The summation
formula may be extended to a class of functions having purely
imaginary poles and satisfying the condition (\ref{Impolecond}).
For this, on the right-hand side of (\ref{SumFormula}) we have to
add the sum of residues (\ref{Impoles2}) and take the principal
value of the second integral on the right-hand side. Using formula
(\ref{SumFormula}), the difference between the sum over the zeros
of the combination of the associated Legendre functions and the
corresponding integral is presented in terms of an integral
involving the Legendre associated functions with real values of
the degree plus residue terms. For a large class of functions
$h(z) $ this integral converges exponentially fast and, in
particular, is useful for numerical calculations. The Abel-Plana
summation formula is obtained as a special case of formula
(\ref{SumFormula}) with $\mu =1/2$ and for an analytic function
$h(z)$. Applying the summation formula for the series over the
zeros of the function $X_{iz}^{\mu }(\cosh (\lambda _{u}/s),\cosh
(\lambda _{v}/s))$ and taking the limit $s\rightarrow \infty $, we
have obtained formula (\ref{SumFormBess}) for the summation of the
series over zeros of the combination of the Bessel functions. The
latter is a special case of the formula, previously derived in
\cite{Sah1}.

A physical application of the summation formula is given in section \ref%
{sec:Phys}. For a quantum scalar field with the general curvature
parameter we have evaluated the positive frequency Wightman
function and the vacuum expectation value of the field squared for
the geometry of concentric spherical shells in a constant negative
curvature space. The Dirichlet boundary conditions on both shells
are assumed. In the region between the shells the eigenfunctions
have the form (\ref{eigfunc2}) and the corresponding
eigenfrequencies are related to the zeros of the function
$X_{iz}^{l+1/2}(u_{1},u_{2})$ by the formula (\ref{eigfreq}). For
the evaluation of the corresponding series in the mode-sum
(\ref{WF1}) for the Wightman function we apply summation formula
(\ref{SumFormula}) with the function $h(z)$ given by (\ref{hz}).
As a result this function is presented in the decomposed form
(\ref{WF2}), where the first term on the right is the Wightman
function for the region outside a single spherical boundary and
the second one is induced by the presence of the outer sphere. By
making use of the identity (\ref{Ident1}), we have presented the
single shell Wightman function as a sum of two terms, formula
(\ref{WF1ext}). The first one is the corresponding function in the
constant curvature space without boundaries and the second one is
induced by the shell. For points away from the shell the latter is
finite in the coincidence limit and can be directly used for the
evaluation of the boundary-induced part in the vacuum expectation
value of the field squared. The renormalization is necessary for
the boundary-free part only and this procedure is the same as that
in quantum field theory without boundaries. In the region between
the spherical shells the vacuum
expectation value of the field squared is presented in the form (\ref{phi2}%
), where the first term on the right-hand side is the corresponding quantity
outside a spherical boundary and is given by the expression (\ref{phi2single}%
).

\section*{Acknowledgements}

The work was supported by the Armenian Ministry of Education and Science
Grant No. 119.

\appendix

\section{On the zeros of the function $X_{iz}^{\protect\mu }(u,v)$}

\label{sec:Zeros}

In this appendix we show that the zeros $z=z_{k}$ are simple and real.
First, we note that the functions $P_{iz-1/2}^{\pm \mu }(u)$ satisfy the
Legendre equation with $\nu =iz-1/2$ and, hence, the function $X_{iz}^{\mu
}(u,v)$ is a solution of the Legendre equation for the same value $\nu $
with respect to both arguments. As a result, by making use of the
differential equation for the associated Legendre functions it can be seen
that the following integration formula takes place%
\begin{equation}
\int_{u}^{v}du\,X_{\nu ^{\prime }}^{\mu }(u,v)X_{\nu }^{\mu }(u,v)=(1-u^{2})%
\frac{X_{\nu ^{\prime }}^{\mu }(u,v)\partial _{u}X_{\nu }^{\mu }(u,v)-X_{\nu
}^{\mu }(u,v)\partial _{u}X_{\nu ^{\prime }}^{\mu }(u,v)}{\nu ^{\prime
2}-\nu ^{2}}.  \label{int1n}
\end{equation}%
Taking the limit $\nu ^{\prime }\rightarrow \nu $ and applying the
l'H\^{o}pital's rule for the right-hand side, from this formula we
find
\begin{equation}
\int_{u}^{v}du\,[X_{iz}^{\mu }(u,v)]^{2}=-\frac{u^{2}-1}{2z}\left\{
[\partial _{z}X_{iz}^{\mu }(u,v)]\partial _{u}X_{iz}^{\mu }(u,v)-X_{iz}^{\mu
}(u,v)\partial _{z}\partial _{u}X_{iz}^{\mu }(u,v)\right\} .  \label{int3}
\end{equation}%
By taking into account the relation $X_{-iz}^{\mu
}(u,v)=X_{iz}^{\mu }(u,v)$, we see that for real $z$ one has
$[X_{iz}^{\mu }(u)]^{2}=|X_{iz}^{\mu }(u)|^{2}$ and the integral
on the left-hand
side of (\ref{int3}) is positive. Now from (\ref{int3}) it follows that $%
[\partial _{z}X_{iz}^{\mu }(u,v)]_{z=z_{k}}\neq 0$, and, hence, the zeros $%
z_{k}$ are simple.

Now let us show that all zeros of the function $X_{iz}^{\mu }(u,v)$ are
real. Suppose that $z=\lambda $ is a zero of $X_{iz}^{\mu }(u,v)$ which is
not real. As the function $X_{iz}^{\mu }(u,v)$ has no pure imaginary zeros, $%
\lambda $ is not a pure imaginary. If $\lambda ^{\ast }$ is the complex
conjugate to $\lambda $, then it is also a zero of $X_{iz}^{\mu }(u,v)$,
because $X_{i\lambda ^{\ast }}^{\mu }(u,v)=[X_{i\lambda }^{\mu }(u,v)]^{\ast
}$. As a result, from formula (\ref{int1n}) we find%
\begin{equation}
\int_{u}^{v}dv\,X_{i\lambda ^{\ast }}^{\mu }(u,v)X_{i\lambda }^{\mu }(u,v)=0.
\label{int4}
\end{equation}%
We have obtained a contradiction, since the integrand on the left hand-side
is positive. Hence the number $\lambda $ cannot exist and the function $%
X_{iz}^{\mu }(u)$ has no zeros which are not real.


\begin{thebibliography}{99}

\bibitem{Grib94} A.A. Grib, S.G. Mamayev, and V.M. Mostepanenko, \textit{%
Vacuum Quantum Effects in Strong Fields} (Friedmann Laboratory Publishing,
St. Petersburg, 1994).

\bibitem{Most97} V.M. Mostepanenko and N.N. Trunov, \textit{The Casimir
Effect and Its Applications} (Oxford University Press, Oxford, 1997).

\bibitem{Birr82} N.D. Birrell and P.C.W. Davis, \textit{Quantum Fields in
Curved Space} (Cambridge University Press, Cambridge, England, 1982).

\bibitem{Noji00} S. Nojiri, S. Odintsov, and S. Zerbini, Class. Quantum
Grav. \textbf{17}, 4855 (2000); W. Naylor and M. Sasaki, Phys. Lett. B
\textbf{542}, 289 (2002); E. Elizalde, S. Nojiri, S.D. Odintsov, and S.
Ogushi, Phys. Rev. D \textbf{67}, 063515 (2003); I.G. Moss, W. Naylor, W.
Santiago-Germ\'{a}n, M. Sasaki, Phys. Rev. D \textbf{67}, 125010 (2003); A.
Flachi, A. Knapman, W. Naylor, and M. Sasaki, Phys. Rev. D \textbf{70},
124011 (2004); J.P. Norman, Phys.Rev. D \textbf{69}, 125015 (2004); W.
Naylor and M. Sasaki, Prog. Theor. Phys. \textbf{113}, 535 (2005); M.
Minamitsuji, W. Naylor, and M. Sasaki, Nucl. Phys. B \textbf{737}, 121
(2006).

\bibitem{Saha08} A.A. Saharian, J. Phys. A: Math. Theor. \textbf{41}, 415203
(2008).

\bibitem{Sah1} A.A. Saharian, Izv. AN Arm. SSR. Matematika \textbf{22}, 166
(1987) [Sov. J. Contemp. Math. Analysis, \textbf{22}, 70 (1987)].

\bibitem{Saha00Rev} A.A. Saharian, "The generalized Abel-Plana formula.
Applications to Bessel functions and Casimir effect," Preprint IC/2000/14;
hep-th/0002239.

\bibitem{Saha07Rev} A.A. Saharian, "The generalized Abel-Plana formula with
applications to Bessel functions and Casimir effect," Preprint
ICTP/2007/082; arXiv: 0708.1187.

\bibitem{Abra72} \textit{Handbook of Mathematical Functions}, edited by M.
Abramowitz and I. A. Stegun (Dover, New York, 1972).

\bibitem{Saha01} A.A. Saharian, Phys. Rev. D \textbf{63}, 125007 (2001).

\bibitem{Saha04} A.A. Saharian and M.R. Setare, Int. J. Mod. Phys. A \textbf{%
19}, 4301 (2004); A.A. Saharian, Nucl. Phys. B \textbf{712}, 196
(2005); A.A. Saharian, Phys. Rev. D \textbf{73}, 044012 (2006);
A.A. Saharian, Phys. Rev. D \textbf{73}, 064019 (2006); E.R.
Bezerra de Mello and A.A. Saharian, Class. Quantum Grav.
\textbf{23}, 4673 (2006); A.A. Saharian and A.S. Tarloyan, J.
Phys. A: Math. Gen \textbf{39}, 13371 (2006); A.A. Saharian and
A.S. Tarloyan, Ann. Phys. \textbf{323}, 1588 (2008).

\bibitem{Eliz94} E. Elizalde, S.D. Odintsov, A. Romeo, A.A. Bytsenko, and S.
Zerbini, \textit{Zeta regularization techniques with applications}
(World Scientific, Singapore, 1994); E. Elizalde, S. Leseduarte,
and A. Romeo, J. Phys. A \textbf{26}, 2409 (1993); S. Leseduarte
and A. Romeo, J. Phys. A \textbf{27}, 2483 (1994); M. Bordag, J.
Phys. A \textbf{28}, 755 (1995); M. Bordag and K. Kirsten, Phys.
Rev. D \textbf{53}, 5753 (1996); M. Bordag, E. Elizalde, and K.
Kirsten, J. Math. Phys. \textbf{37}, 895 (1996); S. Leseduarte and
A. Romeo, Ann. Phys. \textbf{250}, 448 (1996); M. Bordag, K.
Kirsten, and J.S. Dowker, Commun. Math. Phys. \textbf{182}, 371
(1996); M. Bordag, E. Elizalde, K. Kirsten, and S. Leseduarte,
Phys. Rev. D \textbf{56}, 4896 (1997); E. Elizalde, M. Bordag, and
K. Kirsten, J. Phys. A \textbf{31}, 1743 (1998); V.V. Nesterenko
and I.G. Pirozhenko, Phys. Rev. D \textbf{57}, 1284 (1998).
\end{thebibliography}
\end{document}